\documentclass[english]{paper}
\usepackage[T1]{fontenc}
\usepackage[latin9]{inputenc}
\usepackage[a4paper]{geometry}
\usepackage{amsmath}
\usepackage{graphicx}
\usepackage{hyperref}

\makeatletter
\newcommand{\lyxaddress}[1]{
	\par {\raggedright #1
	\vspace{1.4em}
	\noindent\par}
}

\makeatother

\usepackage{babel}
\begin{document}
\title{On the strength of underscreening }
\author{\textup{Andreas Härtel}\textsuperscript{a}\textup{ and Roland Kjellander}\textsuperscript{b}\textup{ }}
\maketitle

\lyxaddress{\emph{a}) Institute of Physics, University of Freiburg, Hermann-Herder-Straße
3, 79104 Freiburg, Germany; \emph{b})\emph{ }Department of Chemistry
and Molecular Biology, University of Gothenburg, P.O. Box 462, SE-405
30 Gothenburg, Sweden. }
\begin{quotation}
This paper constitutes a presentation of work in progress at a discussion
session of the conference \emph{Dense ionic fluids Faraday Discussion,}
8-10 July 2024, London, and is published on pages 293-295 in Ref.
\cite{RKj-2024}.
\end{quotation}
Surface forces between two macroscopic bodies decay for large separations
with the \emph{same decay length as in the bulk phase} in contact
with the surfaces, but the \emph{amplitude} and, for oscillatory forces,
the \emph{phase} depend on the properties of the bodies. 

The Debye length $1/\kappa_{\textrm{DH}}$ for a primitive model electrolyte
solution is obtained from
\[
\kappa_{\textrm{DH}}^{2}=\frac{\beta}{\varepsilon_{0}\varepsilon_{\textrm{r}}}\sum_{j}n_{j}q_{j}^{2}\qquad(\textrm{Debye-Hückel}),
\]
where $\beta=1/k_{B}T$ ($k_{B}$ = Boltzmann's constant, $T$ = absolute
temperature), $\varepsilon_{0}$ = permittivity of vacuum, $\varepsilon_{\textrm{r}}$
= dielectric constant of the dielectric continuum, $q_{j}$ = ionic
charge and $n_{j}$ = number density of species $j$. In the Debye-Hückel
approximation, the mean potential due to a central $i$-ion of diameter
$d$ is given by $\psi_{i}(r)=[q_{i}^{\textrm{eff},\mathrm{DH}}/\varepsilon_{0}\varepsilon_{\textrm{r}}]\cdot e^{-\kappa_{\textrm{DH}}r}/(4\pi r)$
for $r\geq d$, where $q_{i}^{\textrm{eff},\mathrm{DH}}=e^{\kappa_{\textrm{DH}}d}/(1+\kappa_{\textrm{DH}}d)$,
which is an effective charge of this ion. All other ions are treated
as point charges that do correlate with each other in this approximation.

The \textbf{dressed ion theory} (DIT) is an exact reformulations of
the statistical mechanics of simple electrolytes \cite{RKj-1992,RKj-1994}
that has been generalized to ionic liquids and other fluids (see below).
It gives the actual decay length $1/\kappa$ of the electrolyte from
the following exact equation for the decay parameter $\kappa$

\begin{equation}
\kappa^{2}=\frac{\beta}{\varepsilon_{0}\varepsilon_{\textrm{r}}}\sum_{j}n_{j}q_{j}q_{j}^{\textrm{eff}}(\kappa)\qquad(\textrm{spherical ions}),\label{eq:kappa-equation}
\end{equation}
where $q_{j}^{\textrm{eff}}$ is the effective charge of any of the
$j$-ions (not only the central one). Its value differs from $q_{j}^{\textrm{eff},\mathrm{DH}}$.
The \textbf{only difference} between $\kappa_{\textrm{DH}}$ and the
exact $\kappa$ is that one of the factors $q_{j}$ in the equation
is replaced by $q_{j}^{\textrm{eff}}$. This entity appears in the
magnitude of the mean potential $\psi_{j}(r)$ from a $j$-ion, which
decays like
\[
\psi_{j}(r)\;\sim\;\frac{q_{j}^{\textrm{eff}}(\kappa)}{\varepsilon_{0}\mathcal{E}_{\textrm{r}}^{\textrm{eff}}(\kappa)}\cdot\frac{e^{-\kappa r}}{4\pi r}\quad\mathrm{when}\;r\rightarrow\infty,
\]
where $\mathcal{E}_{\textrm{r}}^{\textrm{eff}}(\kappa)$ is an effective
dielectric permittivity that depends on the electrolyte concentration.
It appears here instead for $\varepsilon_{\textrm{r}}$ of the pure
solvent, which is not appropriate for the magnitude of $\psi_{j}$
in an electrolyte. (In eqn \eqref{eq:kappa-equation} for primitive
model electrolytes there is, however, a factor $\varepsilon_{\textrm{r}}$
also in the exact case.)

Eqn \eqref{eq:kappa-equation} has the variable $\kappa$ on both
sides, so it is an equation for $\kappa$ as the unknown entity. It
has multiple solutions $\kappa$, $\kappa'$, $\kappa''$ etc.~that
give the \textbf{decay modes }for interactions in the electrolyte.
Each solution gives rise to a term in the mean potential like
\[
\frac{q_{j}^{\textrm{eff}}(\kappa')}{\varepsilon_{0}\mathcal{E}_{\textrm{r}}^{\textrm{eff}}(\kappa')}\cdot\frac{e^{-\kappa'r}}{4\pi r},\quad\frac{q_{j}^{\textrm{eff}}(\kappa'')}{\varepsilon_{0}\mathcal{E}_{\textrm{r}}^{\textrm{eff}}(\kappa'')}\cdot\frac{e^{-\kappa''r}}{4\pi r},
\]
and so on, each with is own values of $q_{j}^{\textrm{eff}}$ and
$\mathcal{E}_{\textrm{r}}^{\textrm{eff}}$. (Every solution also gives
rise to additional contributions with shorter decay lengths (like
$[2\kappa]^{-1}$, $[3\kappa]^{-1}$, etc.) than the primary term
given above. They are often significantly less important, at least
for large $r$.)

The solutions can be complex-valued, for example, $\kappa'=\kappa'_{\Re}+\mathrm{i}\kappa'_{\Im}$,
$\kappa''=\kappa'_{\Re}-\mathrm{i}\kappa'_{\Im}$, where the latter
is the complex conjugate of the former. Each pair of such solutions
gives rise to an oscillatory contribution. One may for instance have
a monotonic and an oscillatory term
\[
\psi_{j}(r)\;\sim\;C\:\frac{e^{-\kappa r}}{4\pi r}\;+\;C'\,\frac{e^{-\kappa'_{\Re}r}}{2\pi r}\cos(\kappa'_{\Im}r+\vartheta)\quad\mathrm{when}\;r\rightarrow\infty,
\]
where $C$ and $C'$ are constant prefactors obtained from the expressions
above and $\kappa<\kappa'_{\Re}$. 

\begin{figure}[h]
\centering{}\includegraphics[scale=0.8]{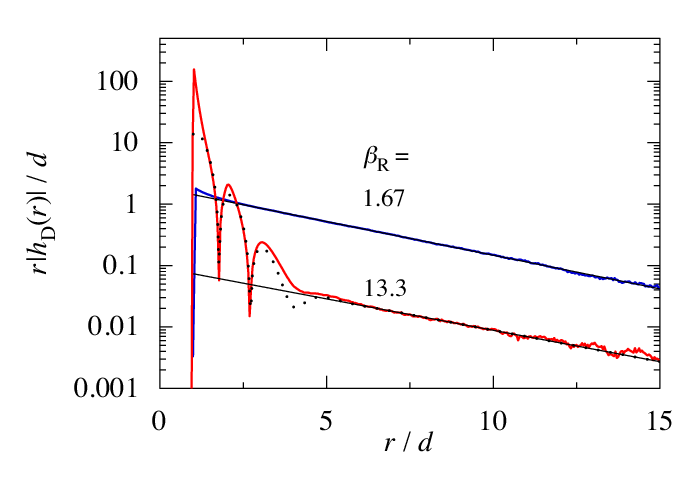}\caption{\label{fig:Function-h_D(r)-plot}The function $h_{\mathrm{D}}(r)$
calculated by Molecular Dynamic (MD) simulations for a dilute electrolyte
solution (0.1 M) in the restricted primitive model for various values
of $\beta_{\textrm{R}}$. \cite{Haertel-2023} It is plotted as $|rh_{\mathrm{D}}(r)|$
on a semi-logarithmic scale. The value $\beta_{\textrm{R}}=1.67$
corresponds to an aqueous solution of monovalent electrolyte at room
temperature. The straight lines show the monotonic tail of each curve.
The dotted curve for $\beta_{\textrm{R}}=13.3$ shows an approximate
fit to eqn \eqref{eq:h_D-asympt}. }
\end{figure}

For a symmetric electrolyte of ions with diameter $d$, the charge-charge
correlation function $H_{\mathrm{QQ}}(r)\,\propto\,h_{\mathrm{D}}(r)\,\equiv\,{\textstyle \frac{1}{2}}[h_{++}(r)-h_{+-}(r)]$.
The function $h_{\mathrm{D}}(r)$ is plotted in Fig. \ref{fig:Function-h_D(r)-plot}
for a dilute electrolyte solution at two different values of the electrostatic
coupling strength $\beta_{\textrm{R}}=\beta q^{2}/[4\pi\varepsilon_{0}\varepsilon_{\textrm{r}}d]$,
where $q$ is the absolute value of the ionic charge. The result for
$\beta_{\textrm{R}}=1.67$ is dominated by a monotonic mode with a
$\kappa$ value not too different from $\kappa_{\textrm{DH}}$, which
is typical for dilute solutions at low coupling. The curve for $\beta_{\textrm{R}}=13.3$
is, however, strikingly similar to surface force curves for many dense
ionic liquids, for instance those shown in Timothy S. Grove's work
presented at this \emph{Faraday Discussions}.\cite{Groves-2024} It
has a short-range oscillatory part and a monotonic tail with a very
long decay length compared to the Debye length ($\ensuremath{\kappa\ll\kappa_{\textrm{DH}}}$).
Like for ionic liquids, the magnitude of the tail is very small and
we will investigate why.

The function $h_{\mathrm{D}}(r)$ and hence $H_{\mathrm{QQ}}(r)$
decay like 
\begin{equation}
h_{\mathrm{D}}(r)\;\sim\;K\:\frac{e^{-\kappa r}}{4\pi r}\;+\;K'\,\frac{e^{-\kappa'_{\Re}r}}{2\pi r}\cos(\kappa'_{\Im}r+\gamma(\kappa'))\quad\mathrm{when}\;r\rightarrow\infty,\label{eq:h_D-asympt}
\end{equation}
where $K$ and $K'$ are constant prefactors and $\gamma$ is the
phase. Eqn \eqref{eq:kappa-equation} for $\kappa$ can in this case
be written $(\kappa/\kappa_{\textrm{DH}})^{2}=q^{\textrm{eff}}(\kappa)/q$,
which means that underscreening (i.e., $\ensuremath{\kappa<\kappa_{\textrm{DH}}}$)
occurs when $q^{\textrm{eff}}(\kappa)<q$. This is the \emph{precise
and unique condition for underscreening to happen} in the present
case. For large underscreening, where $\ensuremath{\kappa\ll\kappa_{\textrm{DH}}}$,
the effective charge is just a small fraction of the bare ionic charge
$q$. 

The prefactor $K$ is proportional to $[q^{\textrm{eff}}(\kappa)]^{2}/\mathcal{E}_{r}^{\mathrm{eff}}(\kappa)$,
where the square of $q^{\textrm{eff}}$ occurs because $h_{\mathrm{D}}$
is a pair function. Thus, $K$ is proportional to $\kappa^{4}/\mathcal{E}_{r}^{\mathrm{eff}}(\kappa)$.
The corresponding applies to $K'$, so we have 
\begin{equation}
\frac{K}{K'}=\left[\frac{\kappa}{|\kappa'|}\right]^{4}\cdot\frac{|\mathcal{E}_{r}^{\mathrm{eff}}(\kappa')|}{|\mathcal{E}_{r}^{\mathrm{eff}}(\kappa)|},\label{eq:K-ratio}
\end{equation}
where $\kappa<|\kappa'|$ in the present case. For large underscreening,
where $\kappa$ is small, $\kappa^{4}$ is much smaller. In the MD
calculations of Fig \ref{fig:Function-h_D(r)-plot} we have $[\kappa/|\kappa'|]^{4}\approx10^{-5}$
when $\beta_{\textrm{R}}$ lies between 13 to 17 and then $K/K'$
from the simulations is about $10^{-3}$ to $10^{-4}$. Thus, the
main reason why the monotonic tail has a \textbf{much smaller magnitude}
than the oscillatory part is\textbf{ the long decay length} of the
former. In the general case, it may therefore be difficult to detect.

DIT has been generalized to arbitrary electrolytes with molecular
solvent, ionic liquids and polar fluids \cite{RKj-2019} and eqn \eqref{eq:K-ratio}
holds for $H_{\mathrm{QQ}}(r)$ also for electrolytes with molecular
solvent and for dense ionic liquids. Thus, the arguments about the
small monotonic long-range tail in $H_{\mathrm{QQ}}(r)$ are valid
in general.


\begin{thebibliography}{1}
\bibitem{RKj-2024}A. P. Abbott, R. Atkin, D. W. Bruce, P. Carbone,
G. Damilano, R. A. W. Dryfe, J.-F. Dufr\^{e}che, K. J. Edler, Y.
K. C. Fung, K. Goloviznina, M. Costa Gomes, A. Grimaud, T. S. Groves,
J. M. Hartley, J. D. Holbrey, C. Holm, P. Illien, R. Kjellander, A.
Kornyshev, K. R. J. Lovelock, D. M. Markiewitz, J. Maurer, S. Miao,
N. Nishi, B. Rocha de Moraes, B. Roling, B. Rotenberg, J. Sangoro,
N. Schaeffer, M. Schönhoff, D. J. Sconyers, J. M. Slattery, M. Swad\'{z}ba-Kwa\'{s}ny,
A. van den Bruinhorst and T. Welton, \href{https://doi.org/10.1039/D4FD90035A}{\emph{Faraday Discuss.} 2024,
\textbf{253}, 289}.

\bibitem{RKj-1992}R. Kjellander and D. J. Mitchell, \href{https://doi.org/10.1016/0009-2614(92)87048-T}{\emph{Chem. Phys.
Lett.,} 1992, \textbf{200,} 76}.

\bibitem{RKj-1994}R. Kjellander and D. J. Mitchell, \href{https://doi.org/10.1063/1.468116}{\emph{J. Chem.
Phys.,} 1994\textbf{, 101,} 603}.

\bibitem{Groves-2024}T. S. Groves and S. Perkin, \href{https://doi.org/10.1039/D4FD00040D}{\emph{Faraday Discuss.,}
2024, \textbf{253}, 193}.

\bibitem{Haertel-2023}A. H\"{a}rtel, M. B\"{u}ltmann and F. Coupette,
\href{https://doi.org/10.1103/PhysRevLett.130.108202}{\emph{Phys. Rev. Lett.}, 2023, \textbf{130}, 108202}.

\bibitem{RKj-2019}R. Kjellander, \href{https://doi.org/10.1039/C9SM00712A}{\emph{Soft Matter}, 2019, \textbf{15,}
5866}.

\end{thebibliography}
\end{document}